\documentclass[9pt,conference]{IEEEtran}
\usepackage{dcase2026}          % single-blind: author names are shown by default
\usepackage{bm}
\usepackage{tabularx}
\usepackage{multirow}
\usepackage{tikz}
\usepackage{booktabs}   % \toprule, \midrule, \bottomrule
\usepackage{tabularx}   % tabularx 환경, X 컬럼, \columnwidth
\usetikzlibrary{arrows.meta,positioning,fit,calc}
\usepackage{pifont}
\newcommand{\cmark}{\ding{51}}   % amssymb를 쓰면 \newcommand{\cmark}{\checkmark}

\title{Parameter Isolation with Domain-Specific Experts for Incremental Audio Classification}

\twoauthors
  {\begin{tabular}[t]{@{}c@{}}
     Jongyeon Park$^{1}$, Do-Hyeon Lim$^{1}$,\\
     Sang-won Park$^{2}$, Hong Kook Kim$^{1,2,3}$%
     \thanks{This work was partly supported by Hanwha Vision Co. Ltd., by the
     Technology Innovation Program (RS-2025-25454727) funded by Korea Ministry of
     Trade, Industry \& Energy, and by the `Project for science and technology opens
     the future of the region' program, funded by the Ministry of Science and ICT
     (MSIT) and Gwangju Metropolitan City, Republic of Korea in 2026.}\end{tabular}}
  {$^{1}$Dept. of AI Convergence, $^{2}$Dept. of EECS\\
   Gwangju Institute of Science and Technology (GIST),\\
   $^{3}$AunionAI Co., Ltd., Gwangju, Korea\\
   \{jypark3737, do-hyeon, sangwonpark\}@gm.gist.ac.kr\\
   hongkook@gist.ac.kr}
  {\begin{tabular}[t]{@{}c@{}}
     Kyungdeuk Ko$^{4}$, Hyeongcheol Geum$^{4}$,\\
     Jeong Eun Lim$^{4}$
   \end{tabular}}
  {$^{4}$AI Lab., R\&D Center, Hanwha Vision\\
   Seongnam, Korea\\
   \{kd.kyung, hch.geum, je04.lim\}@hanwha.com}

\begin{document}

\maketitle

\begin{abstract}
To successfully deploy a model in time-varying environments such as streaming data prediction and sensing control, domain-incremental learning (DIL) has attracted attention since it aims to adapt a previously trained model to newly arriving domains, while preserving knowledge from earlier domains without accessing their data. Incremental learning across domains can be regarded as a recurrent update, in which the current model is obtained by updating the model carried over from previous domains. Conventional DIL approaches that rely on domain-invariant feature learning and weight regularization gradually overwrite or constrain parameters learned in previous domains, leading to catastrophic forgetting. Instead, this paper proposes a new domain-specific parameter-isolation architecture that retains all past domains. 
The proposed architecture mitigates catastrophic forgetting through a full-order recurrent update, constructing a new expert using domain-specific data conditioned on all previously frozen models. To achieve this, we incorporate data-free generative replay to reconstruct previous-domain data and cross-domain feature generation to recover later expert features missing from earlier domain samples. Finally, we apply the proposed model architecture to domain-agnostic incremental learning for audio classification, as defined in the DCASE 2026 Challenge Task 7. Consequently, we achieve micro and macro accuracies of 78.4\% and 78.9\%, respectively, representing increases of 33 and 25 percentage points over the Challenge baseline. Ablation studies are conducted to examine the effectiveness of each processing component in terms of classification accuracy.
\end{abstract}


% ==========================================================================
\section{Introduction}
\label{sec:intro}

Machine learning models deployed in real-world environments inevitably encounter data whose conditions shift over time, requiring continual adaptation. In audio applications, room acoustics, sound characteristics, and even the type of audio itself can change over time. Such time-varying factors degrade the performance of a deployed model, because it must handle data unseen during training, a phenomenon known as domain shift. To address this, domain-incremental learning (DIL) has attracted increasing attention, as it aims to adapt a previously trained model to newly arriving domains while preserving knowledge from earlier ones without accessing their raw data~\cite{cl-review}.

Existing DIL methods can generally be classified into three categories:
regularization-based, replay-based, and parameter-isolation methods~\cite{cl-review}.
Regularization-based methods, such as elastic weight consolidation (EWC)~\cite{kirkpatrick2017ewc},
add a penalty term that discourages changes to weights important for previous
domains, preserving prior knowledge without storing any past data. However, they
protect this knowledge only indirectly through the penalty, and their performance
depends on a stability-plasticity balance that careful tuning alone cannot fully
resolve~\cite{delange2022clsurvey}. This issue becomes more pronounced over long
domain sequences, as the estimated importance becomes increasingly
unreliable~\cite{schwarz2018progress}. On the other hand, replay-based methods take a different
approach, keeping a small buffer of past exemplars and interleaving them with new
data during training~\cite{rebuffi2017icarl, rolnick2019experience}. This tends to
retain prior knowledge more effectively, but the buffer grows with the number of
domains~\cite{cl-review}.

Parameter-isolation methods avoid both problems by assigning each domain its own parameters and freezing those of earlier domains, thereby preserving prior knowledge directly, without a buffer or penalty. A representative example designed for audio is Audio DIL~\cite{mulimani2025dil}, which divides the network into a domain-shared backbone and a domain-specific processing block, updating only the latter for each new domain.
However, Audio DIL's performance is limited by its shared backbone, which learns features only from the first domain and then remains frozen. Consequently, every subsequent domain is constrained to the initial feature set, which is never updated with new data, a limitation that becomes more pronounced as a domain diverges further from the first.

To address this, we propose full-order incremental learning (FoIL). This new DIL framework reinterprets DIL as a recurrent update process in which knowledge accumulated from previous domains is carried forward and integrated as new domains arrive. It is termed full-order because each update conditions on the entire history of previously trained models, rather than only the most recent or the initial one. Concretely, we assign a dedicated expert to each domain, so that each domain is represented by features learned from its own data rather than a backbone fixed to the first domain. New experts are appended to the accumulated ones rather than replacing them. Since every past expert stays frozen and intact, the combined model never degrades on domains it has already learned. However, realizing this recurrent update under the DIL setting is not straightforward: once a new expert is trained, the raw audio of previous domains is no longer available, and an expert added at a later stage produces no output for samples from earlier domains, leaving gaps in what the shared classifier receives. Thus, we implement the recurrent update through two components that together carry previous-domain knowledge into the current step. Data-free generative replay reconstructs data from the previous domain using the frozen experts without storing any raw audio~\cite{yin2020deepinversion}. At the same time, a cross-domain feature-generation module synthesizes missing expert features, ensuring that the classifier always receives a complete input.

Finally, we apply the proposed framework to domain-agnostic incremental learning for audio classification, as defined in the DCASE 2026 Challenge Task 7~\cite{casciotti2026dcasetask7}. Consequently, the proposed framework achieves micro and macro accuracy of 78.4\% and 78.9\%, corresponding to increases of 33 and 25 percentage points over the Challenge baseline, respectively. We further analyze each component through ablation studies on DCASE 2026 Task 7.

Our main contributions are summarized as follows:
\begin{itemize}
\item We reinterpret DIL as a recurrent update process and propose FoIL, a framework that appends a dedicated expert for each domain, thereby guaranteeing no forgetting of previously learned domains.
\item We make FoIL realizable under the strict DIL setting through data-free generative replay, which reconstructs previous-domain data without storing raw audio, and cross-domain feature generation, which completes the missing expert features.
\item We achieve the best performance among all submissions to DCASE 2026 Challenge Task 7 (78.4\% micro, 78.9\% macro accuracy), with gains of 33 and 25 percentage points over the baseline.
\end{itemize}

Following this Introduction, \Cref{sec:related} reviews related work on parameter-isolation-based DIL. \Cref{sec:method} then presents the proposed FoIL framework for audio classification. \Cref{sec:exp} evaluates the framework on the DCASE 2026 Challenge Task 7 and conducts ablation studies to examine how each component contributes to classification accuracy. Finally, \Cref{sec:concl} concludes the paper. 

% ==========================================================================
\section{Related Work}
\label{sec:related}

DIL for audio classification was recently formalized~\cite{mulimani2025dil} and posed
as the domain-incremental DCASE 2026 Challenge Task 7~\cite{casciotti2026dcasetask7}. As mentioned in \cref{sec:intro}, our proposed framework is based on parameter isolation, generative replay, and cross-domain feature generation. Therefore, we briefly review related work on each of these topics.
 
Parameter isolation mitigates forgetting by allocating domain-specific parameters and later combining them~\cite{rusu2016progressive,mallya2018packnet,domainexpert}.
Closest to our design, the dynamically expandable representation (DER) freezes the previously learned representation. It augments it with a
new feature extractor at each step, concatenating their features for a single
classifier~\cite{yan2021der}.
Adapter-based variants instead isolate only low-rank updates on a frozen backbone
via low-rank adaptation (LoRA)~\cite{hu2022lora}, as in orthogonal knowledge-distilled LoRA (OR-KDL) for DCASE 2026 Task 7~\cite{kim2026orkdl}.

Rehearsal is among the most effective methods for continual learning~\cite{cl-review}.
Exemplar-based replay stores and interleaves past samples with new data~\cite{rebuffi2017icarl,rolnick2019experience}.
When storing raw data is disallowed, generative replay synthesizes data instead~\cite{shin2017replay}.
In our case, we adopt data-free DeepInversion~\cite{yin2020deepinversion}, which inverts a frozen classifier using its batch-normalization statistics, allowing each new expert to inherit earlier domain knowledge without raw audio.
 
Finally, cross-domain feature generation relates to feature imputation and
hallucination, which synthesizes unobserved features from the available
ones~\cite{hariharan2017lowshot,xian2018fgn,ma2021smil}. Unlike these methods, our approach generates the new expert's missing features from the previous experts' features, giving our classifier a richer representation.

% ==========================================================================
\section{Proposed FoIL}
\label{sec:method}

\begin{figure}[t]
  \centering
  \begin{tikzpicture}[
    >=Latex, font=\footnotesize,
    st/.style={draw, circle, minimum size=6.5mm, inner sep=0pt},
    frz/.style={st, fill=black!8},
    fade/.style={st, draw=black!35, text=black!45},
    lbl/.style={font=\footnotesize\itshape}]

    % (a) first-order
    \node[fade] (a1) at (0,0)   {$M_1$};
    \node[fade] (a2) at (1.5,0) {$M_2$};
    \node[st]   (a3) at (3.0,0) {$M_3$};
    \draw[->] (a1)--(a2); \draw[->] (a2)--(a3);
    \node[lbl, anchor=west] at (3.5,0) {$M_3\!=\!F(M_2)$};
    \node at (1.5,-0.6) {(a)};

    % (b) anchored (initial-state) update
    \node[st]   (c1) at (0,-1.4)   {$M_1$};
    \node[fade] (c2) at (1.5,-1.4) {$M_2$};
    \node[st]   (c3) at (3.0,-1.4) {$M_3$};
    \draw[->] (c1) to[bend left=28] (c2);
    \draw[->] (c1) to[bend left=28] (c3);
    \node[lbl, anchor=west] at (3.5,-1.4) {$M_3\!=\!F(M_1)$};
    \node at (1.5,-2.0) {(b)};

    % (c) full-order
    \node[frz] (b1) at (0,-2.9)   {$M_1$};
    \node[frz] (b2) at (1.5,-2.9) {$M_2$};
    \node[frz] (b3) at (3.0,-2.9) {$M_3$};
    \draw[->] (b1)--(b2); \draw[->] (b2)--(b3);
    \draw[->] (b1) to[bend left=28] (b3);
    % \node[draw, rounded corners, dashed, fit=(b1)(b2)(b3), inner sep=2.6mm] {};
    \node[lbl, anchor=west] at (3.5,-2.9) {$M_3\!=\!F(M_1,M_2)$};
    \node at (1.5,-3.7) {(c)};
  \end{tikzpicture}
  \caption{Conceptual diagram for representing three different realizations of domain incremental update:
  (a)~first-order, (b)~anchored, and (c)~our full-order update.}
  \label{fig:concept}
\end{figure}
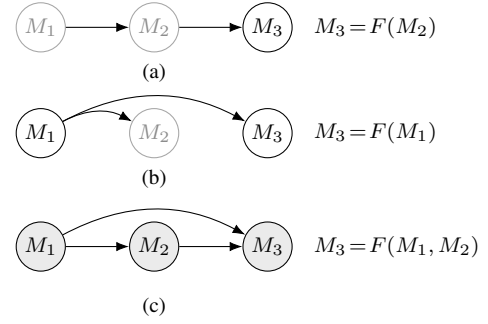

This section presents the FoIL framework for audio classification,
and details the data-free generative replay and cross-domain feature generation
techniques that make it realizable.

\subsection{Motivation of proposed FoIL}
\label{ssec:motivation}
As noted in \cref{sec:intro}, DIL can be regarded as a recurrent update across
domains: the model at the $t$-th domain, $M_t$, is produced from the models
$\{M_i\}_{i=1}^{t-1}$ trained on the earlier domains $\{D_i\}_{i=1}^{t-1}$.
\Cref{fig:concept} compares three realizations of this update for $t=3$.

In the first-order approach (\cref{fig:concept}(a)), $M_t$ is updated only from the
immediately preceding model, i.e., $M_t=F(M_{t-1})$. Regularization- and
distillation-based DIL methods fall under this approach, which is prone to
catastrophic forgetting of the earlier models $M_i$, $i<t-1$. As an alternative
(\cref{fig:concept}(b)), $M_t$ can instead be anchored to the initial model $M_1$,
i.e., $M_t=F(M_1)$. This anchored approach suits settings where every new domain is
invariably a variant of the initial one, such as autonomous driving, hospital-based medical imaging, and voice assistants. Still, its performance degrades under concept
drift, i.e., once a domain departs substantially from the initial domain.

To retain the full history rather than a single reference model, we define the FoIL
framework. When adapting to a new domain, $D_t$, FoIL forms $M_t$ by conditioning on
the entire lineage of previously trained models $\{M_i\}_{i=1}^{t-1}$, not just the
most recent (first-order) or the initial (anchored) one, so that $M_t$ retains the
joint knowledge of every previous domain: $M_t=F(M_1,\dots,M_{t-1})$. For $t=3$, this
gives $M_3=F(M_1,M_2)$, as shown in \cref{fig:concept}(c).

\subsection{Overview of audio classification system}
\label{ssec:overview}

\begin{figure}[t]
  \centering
  \resizebox{0.92\columnwidth}{!}{%
  \begin{tikzpicture}[
      node distance=3mm,
      box/.style={draw, rounded corners=1pt, align=center, font=\scriptsize,
                  minimum height=5mm, inner sep=2.5pt, fill=white},
      frz/.style={box, fill=black!5, minimum width=0.24\columnwidth},
      line/.style={thin}, arr/.style={-{Latex[length=1.4mm]}, thin}]
    \node[font=\scriptsize] (raw) {raw audio of $D_3$};
    \node[box, below=of raw] (mel) {log-mel};
    \node[frz, below=5mm of mel] (e2) {$E_2$\\CNN14 + CRNN};
    \node[frz, left=of e2] (e1) {$E_1$\\CNN14};
    \node[frz, right=of e2] (e3) {$E_3$\\CNN14 + CNN14};
    \node[box, below=5mm of e2, minimum width=0.62\columnwidth] (concat)
        {update function $F$};
    \node[box, below=5mm of concat] (cb)
        {classifier $G_3$};
    \node[font=\scriptsize, below=of cb] (out) {class index};
    \draw[arr] (raw)--(mel);
    \coordinate (tbus) at ($(e2.north)+(0,3mm)$);
    \draw[line] (mel.south) -- (mel.south |- tbus);
    \draw[line] (e1.north |- tbus) -- (e3.north |- tbus);
    \foreach \e in {e1,e2,e3}{\draw[arr] (\e.north |- tbus) -- (\e.north);}
    \foreach \e/\p in {e1/1,e2/2,e3/3}{%
      \draw[arr] (\e.south) -- (\e.south |- concat.north)
        node[midway, anchor=west, font=\scriptsize, inner sep=1pt] {$\phi_\p(D_3)$};}
    \draw[arr] (concat)--(cb) node[midway, anchor=west, font=\scriptsize, inner sep=2pt] {$\mathbf z_3(D_3)$};
    \draw[arr] (cb)--(out);
  \end{tikzpicture}}
  \caption{Domain-agnostic inference pipeline for $T{=}3$. Frozen experts
  $E_1,E_2,E_3$ emit features $\phi_1,\phi_2,\phi_3$, concatenated into $\mathbf z$ and mapped
  to class probabilities $p$ by a single classifier $G$.}
  \label{fig:pipeline}
\end{figure}
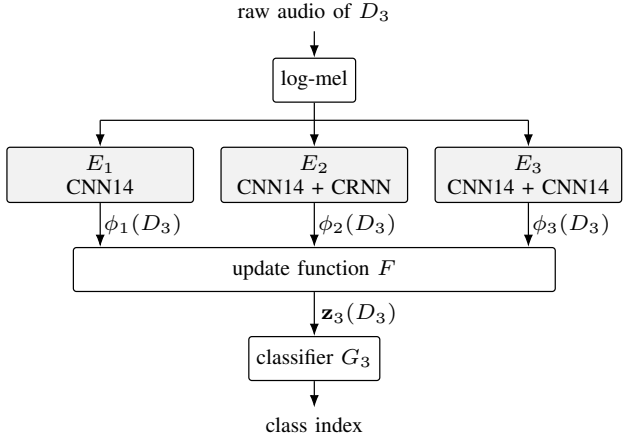

\cref{fig:pipeline} shows the network architecture of the domain-agnostic audio classification inference pipeline, based on the proposed FoIL framework.
First, the input audio is segmented into frames, and log-mel analysis is performed on each frame. Then, the extracted feature vector is applied to each of three experts, $\{E_1, E_2, E_3\}$, resulting in three embedding vectors, $\{\phi_1, \phi_2, \phi_3\}$, which are concatenated into a single vector, $\mathbf z$. The concatenated embedding vector is used as input to the audio classifier to assign a class index to the input frame.

To construct a model in the current domain ($D_t$), $M_t$, we first train an expert, $E_t$, using only data from $D_t$. Specifically, when the audio classification is initiated from the beginning, the initial expert $E_1$ is trained using the data of $D_1$. Thus, the audio classification model $M_1$ is constructed by $M_1$=$E_1$. Next, suppose the domain shifts from $D_1$ to $D_2$. Then, a new expert $E_2$ is trained solely on the data from $D_2$. Consequently, a new audio classification model, $M_2$, is constructed as $M_2 = F(M_1, E_2)$. 
Similarly, for a new domain $D_3$, we repeat the procedure described above, yielding $M_3 = F(M_2, E_3)$. Note that because $M_2$ is a function of $M_1$, it incorporates all prior domain knowledge, aligning with the proposed FoIL framework.

In this paper, all the experts are basically designed using the same neural architecture, CNN14~\cite{kong2020panns}, and the update function, $F(X, Y)$, is realized as concatenating $\mathbf{x}$ and $\mathbf{y}$, which are the output vectors of the experts $X$ and $Y$, respectively. As shown in \cref{fig:pipeline}, $M_i$ is updated using $\mathbf z_i(D_i) =[\mathbf z_{i-1}(D_i), \phi_i(D_i)]$ with $\mathbf z_0(D_i) = \emptyset$ because $M_i = F(M_{i-1}, E_i)$, where $\phi_j(D_i)$ means the embedding vector of $E_j$ using the data of $D_i$.
In practice, the information carried by a new domain can be decomposed into information projected from previous domains and new information specific to the current domain. Therefore, the neural architecture of $E_i$ for $i \geq 2$ needs to combine two sub-networks for projected and new information, respectively. To this end, $E_2$ is extended to include a convolutional recurrent neural network (CRNN), with the former fine-tuned from CNN14 of $E_1$ and the latter trained on $D_2$ data from scratch. Similarly, $E_3$ is also composed of two CNN14s for projected and new information regarding $D_3$.

Lastly, to classify each audio frame, a learnable prototype per class is optimized using a cross-entropy objective~\cite{prototype, wang2017normface}. Then, a cosine similarity-based prototype classifier $G_i$ for the $i$-th domain is applied to each embedding vector $\mathbf z_i (D_i)$.

\subsection{Data-free generative replay}
\label{ssec:inversion}

The proposed FoIL is designed to train two sub-experts: one that retains previous-domain knowledge and another that represents the new domain. As mentioned in the previous subsection, $E_2$ has two networks: CNN14 and CRNN, with CNN14 fine-tuned on $D_2$ data. Therefore, the fine-tuning process does not guarantee that knowledge from the previous domain ($D_1$) is retained in $E_2$ because data from $D_1$ is unavailable.

To remedy this issue, we generate data for $D_1$ using data-free generative replay based on DeepInversion~\cite{yin2020deepinversion}. Specifically, to generate an audio signal belonging to a given class, the previously frozen expert, e.g., $E_1$, is first inferred from random noise with a target class label. The cross-entropy loss is then computed from the resulting output. Simultaneously, the $L_2$ norm is calculated to measure the degree of fit between the mean and variance of the random input and the batch-normalization statistics of the frozen $E_1$. Consequently, the noise input is updated via backpropagation of the sum of the cross-entropy loss and $L_2$ norm.

We then train the new sub-expert, e.g., CNN14 in $E_2$, using a mix of current-domain data and these synthesized samples.
In this way each new expert inherits earlier-domain competence without storing any raw audio, which respects the no-revisit constraint of DIL~\cite{cl-review}.
This expert-training approach, based on data-free generative replay, is one of the main components of the proposed FoIL; it is applied whenever a new domain arrives, although we illustrate it here using $E_1$ and $E_2$ as examples.

\begin{table}[t]
  \centering
  \caption{List of four different systems depending on the configuration of domain experts. Each
  expert is a CNN14, optionally paired with one additional model except for FoIL-FineTune; a subscript of `inv' denotes a
  DeepInversion-based replay variant and `s' means a model trained from scratch.}
  \label{tab:backbones}
  \resizebox{\columnwidth}{!}{%
  \begin{tabular}{@{}llll@{}}
    \toprule
    \multicolumn{1}{@{}c}{System} & \multicolumn{1}{c}{$E_1$ ($D_1$)} & \multicolumn{1}{c}{$E_2$ ($D_2$)} & \multicolumn{1}{c@{}}{$E_3$ ($D_3$)} \\
    \midrule
    FoIL-FineTune& CNN14 & CNN14$_{\text{inv}}$& CNN14$_{\text{inv}}$\\
    FoIL& CNN14 & CNN14$_{\text{inv}}$, CRNN & CNN14$_{\text{inv}}$, CNN14$_{\text{s}}$ \\
    FoIL-FDY& CNN14 & CNN14$_{\text{inv}}$, FDY-CNN14 & CNN14, CNN14$_{\text{s}}$ \\
    FoIL-NoReplay   & CNN14 & CNN14, FDY-CNN14 & CNN14, CNN14$_{\text{s}}$ \\
    \bottomrule
  \end{tabular}}
\end{table}

\subsection{Cross-domain feature generation}
\label{ssec:impute}

Another main component of the proposed FoIL is to utilize the information of $\{E_{j}\}_{j=1}^{i-1}$ for constructing the $i$-th audio classifier, $G_i$, in the feature space.
This is motivated by the observation that the feature space of previous domains overlaps, to some extent, with that of a new domain.
In other words, ${\mathbf z}_{i-1}(D_{i-1})$ could help train $E_i$ if the cross-domain information can be extracted.

Thus, this paper proposes a cross-domain feature generation technique to estimate $\hat{\phi}_{i}(D_{i-1})$ from ${\mathbf z}_{i-1}(D_{i-1})$. To this end, a multi-layer perceptron (MLP)~\cite{rumelhart1986mlp} is trained to approximate this relation, $\hat{\phi}_{i}(D_{i-1})$ = MLP(${\mathbf z}_{i-1}(D_{i-1})$). These two feature embedding vectors are then passed to the update function, $F(\cdot, \cdot)$, to obtain $\mathbf z_i(D_{i-1})$. Finally, a collection of $\mathbf z_i(D_{i})$ and $\mathbf z_i(D_{i-1})$ is used for training $G_i$.

% ==========================================================================
\section{Experiments and Results}
\label{sec:exp}

% ===================== EXPERIMENTS =====================
\subsection{Dataset and evaluation}
\label{ssec:data}

We used the DCASE 2026 Challenge Task 7 DIL dataset, which presented three domains in sequence spanning ten target classes, with the constraint that $D_1$ audio was unavailable, while $D_2$/$D_3$ audio were provided.
Each per-domain test set covered only a subset of the ten classes: $D_2$ has 639 clips (missing \textit{baby\_cry} and \textit{telephone\_ringing}) and $D_3$ has 806 clips (missing \textit{knock}).
Unfortunately, the training set was heavily imbalanced, with counts ranging from $1,125$ clips for \textit{speech} to $170$ for \textit{fire} and $56$ for \textit{baby\_cry}. Since the macro metric was highly sensitive to minority classes, we report micro accuracy (overall clips) and macro accuracy (the mean of per-class accuracies) on the $D_2$ and $D_3$ dev-test sets, along with their average.

\subsection{Implementation details}
\label{ssec:impl}
\Cref{tab:backbones} lists our four systems, depending on the neural network configuration of $E_i$.
FoIL incorporated all the approaches described in Sections \ref{ssec:overview}--\ref{ssec:impute}. FoIL-FDY replaced an auxiliary sub-model of CRNN with frequency dynamic convolution (FDY) for representation diversity. FoIL-FineTune was a version of regular FoIL without auxiliary sub-models, retaining only fine-tuned models. Additionally, FoIL-NoReplay was constructed without replay data because replay data could harm neural network performance, depending on domain characteristics, according to the ablation study in \cref{ssec:ablation}.

All the systems shown in the table shared the same front end as follows: $32$\,kHz mono audio cropped to $4$\,s and converted to a $64$-bin log-mel spectrogram ($1,024$-point window, $320$-point hop,
$f\in[50,\ 14{,}000]$\,Hz). Since the domain tag was unavailable at inference, every system ran a domain-agnostic forward pass once per clip, with no test-time augmentation.
Each expert emitted a $2,048$- or $4,096$-dimensional embedding $\phi_i$, depending on
whether it included an auxiliary sub-model for $E_2$ and $E_3$, and the three were concatenated into a
$10,240$-dimensional vector $\mathbf z$ in our three-domain setting. The classifier was trained
on $\mathbf z$ with balanced sampling~\cite{psla} to counter class imbalance, using Adam
(learning rate~$10^{-3}$, batch $64$, $200$ epochs, cosine schedule); its prototypes were
centroid-initialized from the cached class means and its temperature $\tau$ was learned. The cross-domain feature generation technique was implemented by a two-layer MLP (hidden size $4,096$)
trained on the $D_3$ rows with a mean squared error loss for $50$ epochs.

% ===================== RESULTS =====================
\subsection{Results}
\label{ssec:main}
\begin{table}[t]
  \centering
  \caption{Comparison of micro and macro average classification accuracy (\%) between our proposed systems and comparison systems, including the Challenge baseline, for $D_2$ and $D_3$ domains.}
  \label{tab:systems}
  \sisetup{table-format=2.2, detect-weight}
  \resizebox{\columnwidth}{!}{%
  \begin{tabular}{l *{6}{S}}
    \toprule
     & \multicolumn{2}{c}{$D_2$} & \multicolumn{2}{c}{$D_3$}
     & \multicolumn{2}{c}{Dev-test Avg.} \\
    \cmidrule(lr){2-3}\cmidrule(lr){4-5}\cmidrule(lr){6-7}
    System & {micro} & {macro} & {micro} & {macro} & {micro} & {macro} \\
    \midrule
    Challenge Baseline     & 54.77 & 58.95 & 36.23 & 47.34 & 45.50 & 53.15 \\
    OR-KDL                 & {--}  & 79.73 & {--}  & 66.55 & {--}  & 73.14 \\
    % FoIL          & 81.22 & 82.50 & 73.33 & 73.12 & 77.27 & 77.81 \\
    FoIL-FineTune          & 68.86 & 71.94 & 70.10 & 70.15 & 69.48 & 71.05 \\
    FoIL            & 79.97 & 81.88 & 75.19 & 73.97 & 77.58 & 77.92 \\
    FoIL-Ensemble  & 81.69 & 83.62 & 75.06 & 74.22
                           & \bfseries 78.38 & \bfseries 78.92 \\
    \bottomrule
  \end{tabular}}
\end{table}

\Cref{tab:systems} reports our systems, as described in \cref{tab:backbones}, and their ensemble against the DCASE 2026 Challenge Task 7 baseline~\cite{mulimani2025dil} and OR-KDL~\cite{kim2026orkdl}. 
As shown in the first and second rows of the table, the Challenge
baseline achieved 45.5\% and 53.2\% in micro and macro accuracy, respectively, while
OR-KDL, a knowledge distillation approach with domain-specific low-rank adapters, achieved 73.14\% macro accuracy.
Our proposed single system, FoIL, surpassed both, improving macro accuracy by 24.77 and 4.78 percentage points over the baseline and OR-KDL, respectively. This implies that the proposed FoIL framework benefits domain-incremental audio classification by fully preserving previous-domain knowledge in its experts.

Finally, to further improve the classification accuracy, we constructed an ensemble system by combining all four systems described in \cref{tab:backbones}. Consequently, the ensemble system achieved the highest micro- and macro-accuracy for the dev-test set.

\begin{table}[t]
  \centering
  \caption{Ablation results of the macro accuracy according to different combinations of the processing components in the proposed FoIL framework.}
  \label{tab:ablation}
  \begin{tabular*}{\columnwidth}{@{\extracolsep{\fill}}l ccc ccc@{}}
    \toprule
     & \multicolumn{3}{c}{Components} & \multicolumn{3}{c}{Macro Acc.\ (\%)} \\
    \cmidrule(lr){2-4}\cmidrule(lr){5-7}
    Variant & Replay & \shortstack{MLP-Gen} & \shortstack{Feat-Mean} & $D_2$ & $D_3$ & Avg. \\
    \midrule
    FoIL-None  &  &  &  &  72.88 & 72.05 & 72.47\\
    FoIL-R  & \cmark &        &        & 74.98 & 74.19 & 74.59 \\
    FoIL-M  &        & \cmark &        & 79.22 & 74.42 & 76.82 \\
    FoIL-RM & \cmark & \cmark &        & 81.88 & 73.97 & \textbf{77.92} \\
    FoIL-RF  & \cmark &        & \cmark & 65.33 & 75.50 & 70.42 \\
    % FoIL-3 & \cmark & \cmark &        & 71.94 & 70.15 & 71.05 \\
    \bottomrule
  \end{tabular*}
\end{table}

\subsection{Ablation studies}
\label{ssec:ablation}
In this subsection, we examined the effectiveness of each processing component in FoIL on audio classification accuracy. 
\Cref{tab:ablation} shows the results of this ablation study, in which the processing components studied were data-free generative replay in \cref{ssec:inversion} (denoted as Replay) and MLP-based cross-domain feature generation in \cref{ssec:impute} (MLP-Gen). Additionally, for the performance comparison with MLP-Gen, we simply replaced MLP-Gen-based feature-embedding vector with the class-wise mean vector obtained from the new domain, which was denoted as Feat-Mean. In the table, Replay, MLP-Gen, and Feat-Mean are abbreviated as R, M, and F, respectively. For example, FoIL-RF means a model trained by applying Replay and Feat-Mean techniques, while FoIL-RM is identical to FoIL in \cref{tab:systems}.

We first evaluated the performance of a FoIL system without Replay or MLP-Gen. In other words, FoIL-None was implemented as described in \cref{ssec:overview}. Next, we incorporated Replay into FoIL-None (FoIL-R in the table). Compared with FoIL-None, FoIL-R improved macro accuracy for $D_2$ and $D_3$ owing to the increased amount of training data. Instead of Replay, MLP-Gen was applied to construct the classifier, and it was shown that FoIL-M provided higher macro accuracy than FoIL-R in both domains. In particular, the improvement in classification for $D_2$ was significant, indicating that cross-domain feature generation could contribute more to the transfer of domain knowledge than data-free generative replay.
This phenomenon reappeared in FoIL-RM, further improving macro accuracy for $D_2$ but slightly reducing it for $D_3$. There was a trade-off between preventing forgetting and maintaining current-domain knowledge.

\begin{figure}[t]
  \centering
  \includegraphics[width=\columnwidth]{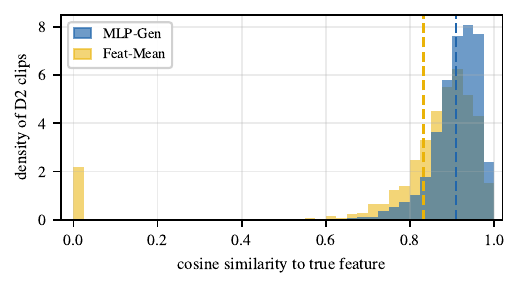}
    \vspace{-0.8cm}
  \caption{Plot of the distribution of cosine similarity scores calculated between the ground truth and generated feature.}
  \label{fig:impute}

\end{figure}
Next, we assessed the quality of the feature generation technique. As shown in the last row of the table, FoIL-RF severely degraded the classification performance for $D_2$ compared with FoIL-RM. This implies that a feature generation technique for incremental learning must be carefully designed. 
To support this with a feature-level comparison between the MLP-Gen and Feat-Mean approaches, we first obtained the ground-truth (GT) feature by applying the data from $D_2$ to an expert $E_3$, $\phi_3(D_2)$. This GT was compared with $\hat{\phi}_3(D_2)$ = MLP(${\mathbf z}_{2}(D_2)$) using cosine similarity. In parallel, we repeated this computation using $\bar{\phi}_3(D_2)$ = Feat-Mean($D_3$). 
\Cref{fig:impute} plots the distribution of cosine similarity scores for MLP-Gen and Feat-Mean, where the long, vertical dotted line indicates the average score for each distribution. Also, the yellow vertical bar around a similarity score of zero was due to a mismatch between the audio class types of $D_2$ and $D_3$.
The distribution is expected to be skewed toward the rightmost end, because the ideal case is an impulse at a similarity of one. As shown in the figure, MLP-Gen provided a better fit to GT than Feat-Mean, resulting in better performance for FoIL-RM than for FoIL-RF.

% ==========================================================================
\section{Conclusion}
\label{sec:concl}

In this paper, we addressed domain-incremental audio classification under the DIL constraint and proposed FoIL, a domain-specific parameter-isolation framework.
Each domain was handled by a dedicated expert that was frozen once trained and appended to the accumulated model, preventing forgetting.
A single classifier over the concatenated features enabled a domain-agnostic forward pass.
Data-free generative replay let each new expert inherit earlier-domain knowledge without stored audio, and an MLP-based cross-domain feature generator completed the expert features for earlier-domain samples.
On DCASE 2026 Challenge Task 7, FoIL improved micro and macro accuracy over the Challenge baseline by $33$ and $25$ percentage points, respectively. The ablation study confirmed the contribution of each component, offering an effective route to audio DIL.

% -------------------------------------------------------------------------
\clearpage
\bibliographystyle{IEEEtran}
\bibliography{refs}

\end{document}